\documentclass[preprint2]{aastex}

\usepackage{rotating}
\usepackage{color}
\usepackage{amsmath}
\usepackage{graphicx}
\usepackage{mwe}
\usepackage{grffile}
\usepackage{multicol}
\usepackage{longtable}
\usepackage{pdflscape}
\usepackage{tabls}
\usepackage{threeparttable}

\begin{document}
\setcounter{secnumdepth}{2}
\title{GPI Spectroscopy of the Mass, Age, and Metallicity Benchmark Brown Dwarf HD~4747~B}
\author{Justin R. Crepp\altaffilmark{1}, David A. Principe\altaffilmark{2,3}, Schuyler Wolff\altaffilmark{4, 5}, Paige A. Giorla Godfrey\altaffilmark{6,7,8}, Emily L. Rice\altaffilmark{6,7,8}, Lucas Cieza\altaffilmark{3}, Laurent Pueyo\altaffilmark{4,9}, Eric B. Bechter\altaffilmark{1}, Erica J. Gonzales\altaffilmark{1}}
\email{jcrepp@nd.edu} 
\altaffiltext{1}{Department of Physics, University of Notre Dame, 225 Nieuwland Science Hall, Notre Dame, IN, 46556, USA}
\altaffiltext{2}{Massachusetts Institute of Technology, Kavli Institute for Astrophysics, Cambridge, MA, USA}
\altaffiltext{3}{N{\'u}cleo de Astronom{\'i}a de la Facultad de Ingenier{\'i}a, Universidad Diego Portales, Av. Ej{\'e}rcito 441, Santiago 8320000, Chile}
\altaffiltext{4}{Space Telescope Science Institute, 3700 San Martin Drive, Baltimore, MD 21218, USA}
\altaffiltext{5}{Leiden Observatory, Leiden University, P.O. Box 9513, 2300 RA Leiden, The Netherlands}
\altaffiltext{6}{College of Staten Island, CUNY, 2800 Victory Boulevard, Staten Island, NY 10314, USA}
\altaffiltext{7}{Physics Program, The Graduate Center, City University of New York, 365 5th Ave, New York, NY 10016, USA}
\altaffiltext{8}{Department of Astrophysics, American Museum of Natural History, 79th and Central Park West, New York, NY 10024}
\altaffiltext{9}{Department of Physics and Astronomy, Johns Hopkins University, Baltimore, MD 21218, USA}

\date{\centerline{Intended for ApJ, Draft Manuscript: \today}}

\begin{abstract}
The physical properties of brown dwarf companions found to orbit nearby, solar-type stars can be benchmarked against independent measures of their mass, age, chemical composition, and other parameters, offering insights into the evolution of substellar objects. The TRENDS high-contrast imaging survey has recently discovered a (mass/age/metallicity) benchmark brown dwarf orbiting the nearby ($d=18.69\pm0.19$ pc), G8V/K0V star HD~4747. We have acquired follow-up spectroscopic measurements of HD~4747~B using the Gemini Planet Imager to study its spectral type, effective temperature, surface gravity, and cloud properties. Observations obtained in the $H$-band and $K_1$-band recover the companion and reveal that it is near the L/T transition (T1$\pm$2). Fitting atmospheric models to the companion spectrum, we find strong evidence for the presence of clouds. However, spectral models cannot satisfactorily fit the complete data set: while the shape of the spectrum can be well-matched in individual filters, a joint fit across the full passband results in discrepancies that are a consequence of the inherent color of the brown dwarf. We also find a $2\sigma$ tension in the companion mass, age, and surface gravity when comparing to evolutionary models. These results highlight the importance of using benchmark objects to study ``secondary effects" such as metallicity, non-equilibrium chemistry, cloud parameters, electron conduction, non-adiabatic cooling, and other subtleties affecting emergent spectra. As a new L/T transition benchmark, HD~4747~B warrants further investigation into the modeling of cloud physics using higher resolution spectroscopy across a broader range of wavelengths, polarimetric observations, and continued Doppler radial velocity and astrometric monitoring.
\end{abstract}

\section{INTRODUCTION}\label{sec:intro}
Precise measurements of the physical properties of stellar and substellar benchmark objects provide falsifiable observational data against which theoretical models may be thoroughly examined and tested \citep{liu_10,dupuy_14}. Of particular value are companions to nearby stars for which mass may be determined through orbital dynamics, and composition and age may be assumed to closely match that of the bright and well-studied host star \citep{crepp_12b}. Mass governs the thermal evolution of substellar objects while composition and cloud properties impacts their emergent spectrum. In the past, it has been difficult to acquire spectra of faint dwarf companions in close angular proximity to nearby stars. However, advances in high-contrast imaging technology have helped to overcome this obstacle by generating sufficient dynamic range to detect the signal of brown dwarfs and giant exoplanets \citep{dohlen_08,hinkley_11,macintosh_15,konopacky_16,bowler_16}. 

Few substellar benchmark objects have been detected for which mass, age, and metallicity are known independently of the light that they emit \citep[e.g.,][]{liu_02,potter_02}. In a recent discovery article from the TRENDS high-contrast imaging program, \citealt{crepp_16} reported the detection of a new benchmark brown dwarf that orbits HD~4747 ($\theta=0\farcs6$, $\rho=11.3\pm0.2$ a.u.). This nearby ($\pi=53.51\pm0.53$ mas), solar-type (G8V/K0V) star has a precisely determined metallicity of [Fe/H]$=-0.22\pm0.04$ and a gyrochronological age of $\tau = 3.3^{+2.3}_{-1.9}$ Gyr. Further, 18 years of precise radial velocity (RV) measurements combined with astrometry from its discovery place strong dynamical constraints on the companion mass demonstrating that it is below the hydrogen fusing limit  ($m\sin i=55.3 \pm 1.9M_{\rm Jup}$). 

We have acquired a spectrum of HD~4747~B using adaptive optics (AO) and the Gemini Planet Imager (GPI) instrument on the Gemini South telescope \citep{macintosh_08,macintosh_14}. In this paper, we study the companion spectral type, effective temperature, surface gravity, and atmospheric properties. We demonstrate that HD~4747~B is an L/T transition object and therefore an interesting benchmark brown dwarf that may be used to test theoretical evolutionary models and spectral models.  

\begin{figure*}[!t]\label{fig:image}
\begin{center}
\includegraphics[height=3.2in]{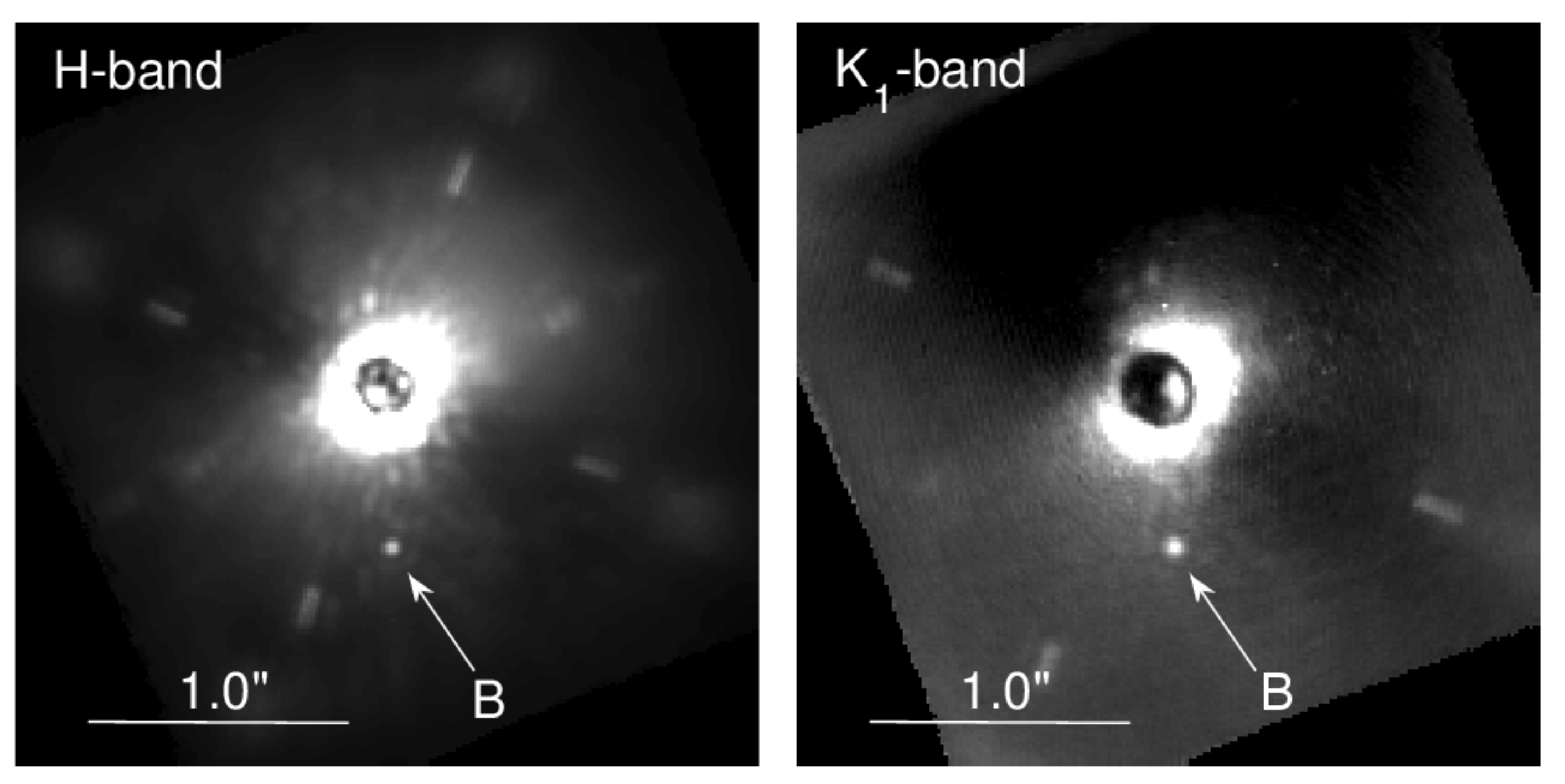} 
\caption{Images of HD~4747~B taken with GPI in the $H$ and $K_1$ filters. The companion is clearly recovered in each band at a high signal-to-noise ratio in wavelength-collapsed frames. North is up and east is to the left. IFS data was also used to acquire a spectrum of HD~4747~B.}
\end{center}
\end{figure*}  

\begin{threeparttable}[!ht]
\begin{tabular}{lcc}
\hline
\multicolumn{3}{c}{HD~4747 Properties}     \\
\hline
Spectral Type                           &       G9V                          &  1, 2   \\
right ascension [J2000]            &   00 49 26.77                   &       \\
declination [J2000]                   &    -23$^{\circ}$ 12 44.93  &       \\
d [pc]                                          &    $18.69\pm0.19$        &  3  \\
proper motion [mas/yr]              &    $516.92\pm0.55$  E  &        \\
                                                   &   $120.05\pm0.45$ N  &         \\
$B$                                           &   7.92   & 4  \\
$V$                                           &   7.16   &  \\
$R$                                           &   6.73   &  \\
$I$                                            &    6.34   &  \\
$J_{\tiny{\rm 2MASS}}$            &    $5.813\pm0.021$ &  5 \\
$H_{\tiny{\rm 2MASS}}$            &   $5.433\pm0.049$ & \\
$K_s$                                        &   $5.305\pm0.029$ &  \\
$L'$                                            &    $5.2\pm0.1$       &   6    \\
\hline
Mass [$M_{\odot}$]           &    $0.82\pm0.04$           &  2  \\
Radius [$R_{\odot}$]         &     $0.79\pm0.03$         &  2  \\
Age [Gyr]                            &     $3.3^{+2.3}_{-1.9}$ & 7 \\
$\mbox{[Fe/H]}$                 &      $-0.22\pm0.04$      & 2  \\
log g [cm $\mbox{s}^{-2}$]  &     $4.55$                 &  8 \\
$T_{\rm eff}$ [$K$]              &    $5340\pm40$         & 2  \\
v sini   [km/s]                        &   $1.1\pm0.5$           &    2    \\
\hline
\end{tabular}
	\begin{tablenotes}
	\small
		\item [1] \citealt{montes_01}.
		\item [2] Spectroscopic Properties of Cool Stars (SPOCS) database \citep{valenti_fischer_05}. 
		\item [3] \citealt{van_leeuwen_07}.
		\item [4] Visible photometry is from \citealt{koen_10}.
		\item [5] NIR photometry is from \citealt{cutri_03}.
		\item [6] L' is estimated from a black-body fit to the stellar spectral energy distribution.
		\item [7] \citet{crepp_16}.
		\item [8] Isochronal estimate from SPOCS database.
	\end{tablenotes}
\caption[]{Properties of the HD~4747 system reproduced from \citet{crepp_16}.} 
\label{tab:starprops}
\end{threeparttable} 

\section{OBSERVATIONS}\label{sec:observations}
HD~4747 was observed on UT 2015 December 24 and 2015 December 25 using the GPI integral field spectrograph (IFS). GPI provides a plate scale of 0.014 arcseconds per pixel and field of view of just under 2$\farcs$8 $\times$ 2$\farcs$8 for imaging as well as low-resolution spectroscopy ($R\approx30-90$) for atmospheric characterization \citep{macintosh_14,macintosh_15}. IFS data is also used to help remove stellar noise from images by taking advantage of the wavelength dependence of speckles \citep{sparks_ford_02,crepp_11,perrin_14}. 

Measurements of HD~4747~B were recorded in coronagraphic mode in the $K_1$ ($\lambda = 1.90 - 2.19 \; \mu$m) and $H$ ($\lambda=1.50 - 1.80 \; \mu$m) bands respectively on each night. A diffractive grid mask located in a pupil plane was used to generate ``satellite spots" that track the position of the star in each channel, facilitating astrometry \citep{konopacky_14}. On the first night (12-24-2015 UT), 38 high-quality cubes were recorded in $K_1$-band resulting in a total on-source integration time of $\Delta t=37.8$ minutes. The airmass ranged from $\sec(z)=1.07-1.16$. Taking the average between MASS and DIMM measurements, zenith seeing was approximately $0\farcs7$. On the second night (12-25-2015 UT), 36 high-quality cubes were recorded in the $H$-band resulting in a total on-source integration time of $\Delta t = 35.8$ minutes. The airmass ranged from $\sec(z)=1.05-1.13$. Seeing conditions were comparable to the first night. Table~\ref{tab:log} displays a log of the observations. On both nights, the amount of parallactic angle rotation ($\Delta \pi$) in each filter was only several degrees. As such, data analysis relied largely on chromatic diversity provided by the integral field unit (IFS) to reduce speckle noise. 

\begin{table}[!ht]
\centerline{
\begin{tabular}{lccc}
\hline
\hline
\multicolumn{4}{c}{Summary of GPI Observations}     \\
\hline
\hline
Date [UT]          & Filter        & $\Delta t$ [min]  & $\Delta \pi$ [$^{\circ}$]  \\
\hline
Dec. 24 2015    & $K_1$      &      37.8              &   4.6      \\ 
Dec. 25 2015    & $H$          &      35.8              &   5.5      \\
\hline
\hline
\end{tabular}}
\caption{GPI observing log showing filter selection, on-source integration time ($\Delta t$), and change in parallactic angle ($\Delta \pi$).}
\label{tab:log}
\end{table} 

\begin{figure*}[th]
\centering\begin{tabular}{c}
\includegraphics[scale=0.43]{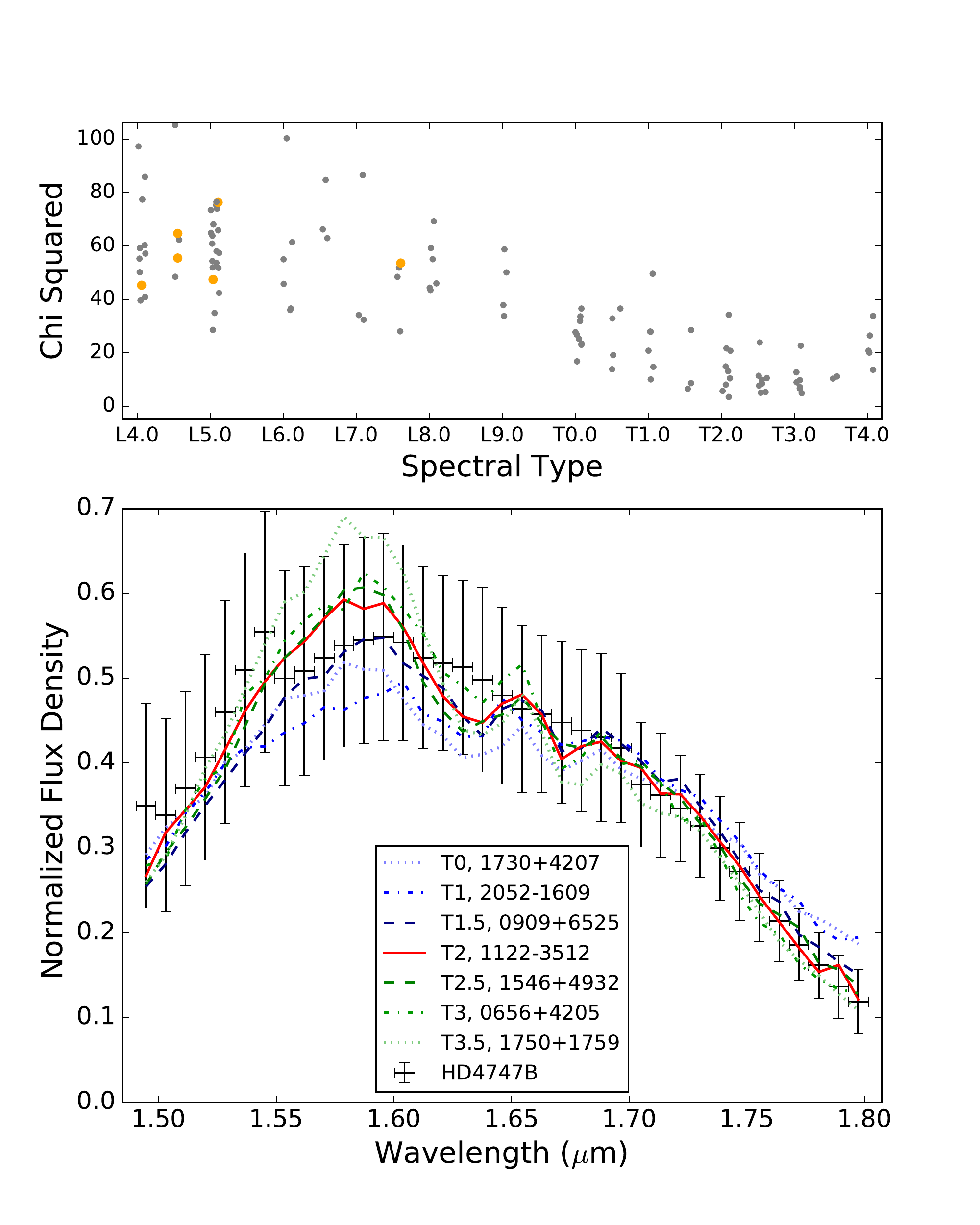} 
\hspace{-0.3in}
\includegraphics[scale=0.43]{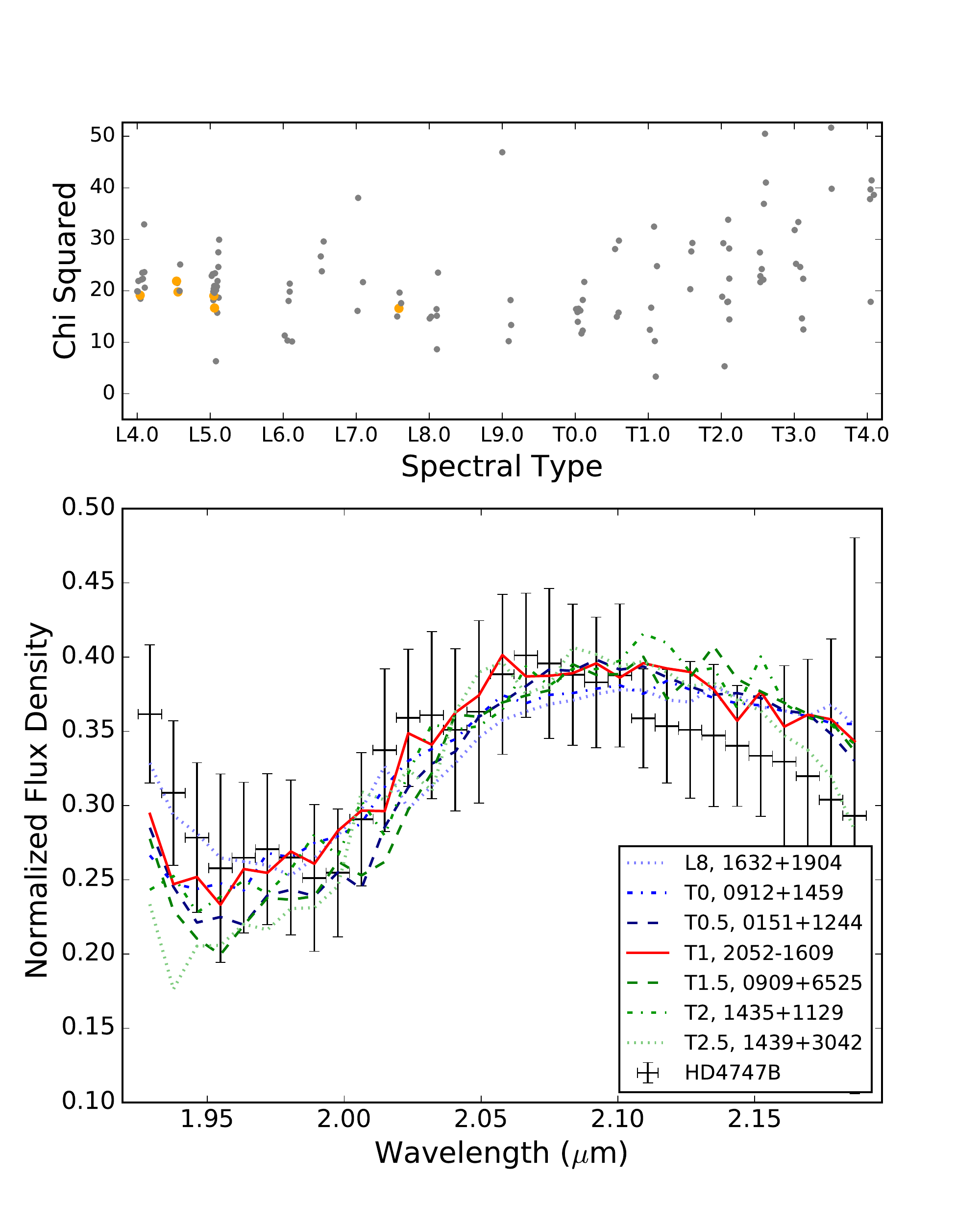} 
\end{tabular}
\caption{Results from $\chi^2$-based spectral type analysis. Goodness of fit values as a function of spectral type for each source are shown in the upper panel (67 degrees of freedom). Gray dots represent old field objects while orange dots are young. The spectral types have been slightly ``jittered" horizontally for visual clarity. The H-band (left) and $K_1$-band (right) are first analyzed separately and then combined (see Fig.~\ref{fig:sptypeboth}).}
\label{fig:sptype}
\end{figure*}

\begin{figure*}[th]
\centering\begin{tabular}{c}
\includegraphics[scale=0.43]{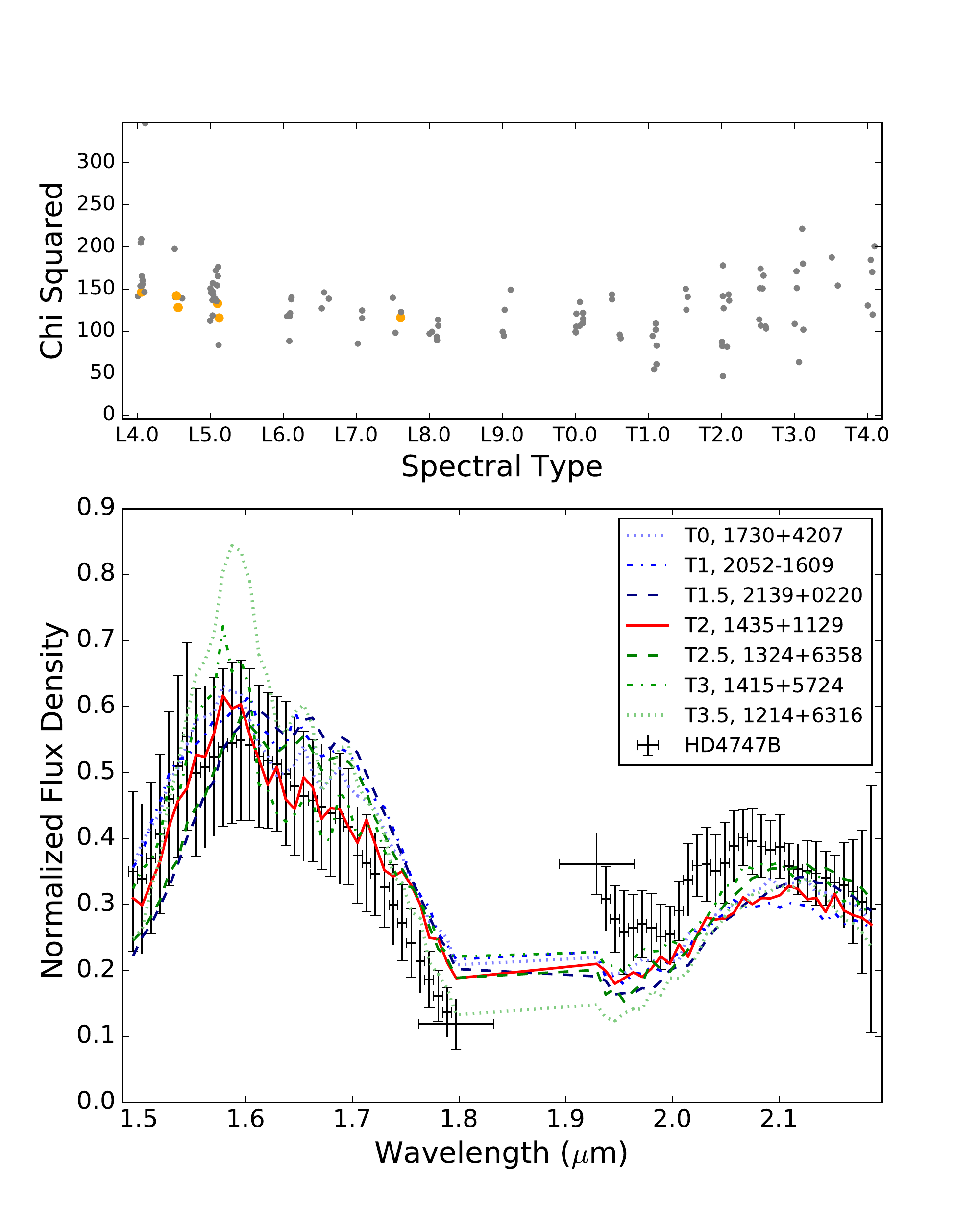}
\end{tabular}
\caption{Same as in Figure~\ref{fig:sptype} but showing both filter data sets fit simultaneously. HD~4747~B appears to be an L/T transition object.}
\label{fig:sptypeboth}
\end{figure*}

\section{DATA REDUCTION}\label{sec:data}
GPI's IFS lenslet array produces $190 \times 190$ individual spectra from which spatial and chromatic information can be extracted \citep{macintosh_14}. Raw frames were converted into data cubes using version 1.3.0 of the GPI Data Reduction Pipeline \citep[see SPIE paper series described in][]{perrin_14}. Images were dark subtracted, corrected for bad pixels, and destriped to accommodate for variations between CCD readout channels and microphonics. Additionally, thermal background subtraction was performed for the \textit{K1} band dataset. An arc lamp image taken immediately preceding each dataset was used for wavelength calibration of each individual lenslet \citep{wolff_14}. 

Image registration and photometric flux calibration were performed using satellite spot copies of the occulted stellar PSF \citep{wang_14}. From 2015 August 6 through 2016 October 11, the focal plane mask was kept in the \textit{H} band position due to a mechanical issue. Therefore, we used the \textit{H} band star-to-satellite spot flux ratio of  $9.23 \pm 0.06$ magnitudes to calibrate both the \textit{H} and \textit{K1} datasets.\footnote{As shown in \citet{sivaramakrishnan_oppenheimer_06}, the relative flux scaling of satellite spots depends only on wire grid geometry and not wavelength.} Spectrophotometric calibration was performed using the pipeline primitive ``Calibrate Photometric Flux."  This primitive uses the averaged spectrum of the satellite spots, i.e. emission from HD~4747 which experiences the same atmospheric and instrumental transmission functions as the companion, with a provided spectral shape determined from the Pickles Library.\footnote{Pickles library at http://www.stsci.edu/hst/observatory/crds/pickles$\_$atlas.html.} The image was flux calibrated assuming a G8V spectrum and H-band 2MASS flux of $H=5.433$ for the primary star \citep{houk_88,cutri_03}. Reduced frames were rotated with North up. 

Figure~\ref{fig:image} shows coronagraphic images of HD~4747 after median combining data sets for each band. The companion is detected in each filter at a consistent angular separation ($\S$\ref{sec:astrometry}). A preliminary spectral analysis that allowed for qualitative characterization indicated that HD~4747~B may be near the L/T transition. Although the companion is sufficiently bright to be noticed in individual exposures, the resulting signal-to-noise ratio was SNR$\sim$9 per wavelength channel. We used speckle suppression based on spectral differential imaging (SDI) to more accurately interpret the data set by enhancing the SNR.

The routine PyKLIP was used to generate an SDI PSF-subtracted data cube \citep{pueyo_12,pueyo_16}. Synthetic ``planets" were embedded into the calibrated data cubes at evenly spaced azimuthal angles and with angular separations comparable to HD~4747~B ($\pm1 \; \lambda/D$) but far enough away to avoid artificial subtraction. The flux of each synthetic object was iteratively adjusted to closely match that of the real companion following speckle suppression, thereby minimizing non-linear effects with subtraction. The PSF profile of the star was used as the spatial profile of synthetic companions. Wavelength-dependent effective throughput was calculated for each synthetic companion to calibrate the effects of self-subtraction and quantify variance. Variations in the resulting photometry and astrometry were used to determine uncertainties in companion spectral and positional properties, along with uncertainties in flux calibration. Multiple pipelines were used to verify results from KLIP. The results were further checked against changing high-pass filter properties and number of Karhunen-Loeve modes. 

\begin{table*}[!ht]
\centerline{
\begin{tabular}{lccccc}
\hline
\hline
\multicolumn{6}{c}{Best Matching Brown Dwarfs}     \\
\hline
\hline
Band      & Brown Dwarf Name   & NIR SpType & Discovery    &  SpType Reference & SpeX Reference \\
\hline \hline
$HK_1$     & SDSS J1435+1129  & T2.0             & Chiu 2006    & Chiu 2006          & Chiu 2006 \\
$H$            & 2MASS J1122-3512 & T2.0             & Tinney 2005 & Burgasser 2006 & Burgasser 2006 \\
$K_1$        & SDSS J2052-1609   &  T1.0            & Chiu2006     &  Chiu2006          & Chiu2006 \\
\hline
\hline
\end{tabular}}
\caption{Best-fitting comparison objects to HD~4747~B.}
\label{tab:bestmatch}
\end{table*}

\section{RESULTS}\label{sec:results}

\subsection{Spectral Typing}\label{sec:sptype}
Late-type dwarfs are assigned a spectral type by comparison of near-infrared spectra to that of spectral standards \citep{burgasser_06a}. We employed this method in Crepp et al. (2015), and use it again here for the spectrum of HD~4747~B. We compare the spectrum of HD 4747~B to 308 L and T dwarf spectra from the SpeX prism library, supplemented by spectra from Filippazzo et al. (2015). Most spectra are from Burgasser et al. (2004, 2006a, 2008, 2010), Burgasser (2007), Chiu et al. (2006), Cruz et al.  (2004), Kirkpatrick et al. (2011), Liebert \& Burgasser (2007), Looper, Kirkpatrick, \& Burgasser (2007), Mace et al. (2013), Mainzer et al. (2011), and Sheppard \& Cushing (2009).

We begin by binning and trimming comparison spectra to match the resolution and wavelength coverage of the HD 4747~B GPI spectrum. We compare the L and T dwarf spectra to that of HD~4747~B with a normalization constant and $\chi^2$-like goodness of fit statistic from Cushing et al. (2008). The $\chi^2$ value is calculated for each binned L and T dwarf spectrum, incorporating errors from spectral templates and the spectrum of HD~4747~B. We perform this fit to the $H$ and $K_1$ bands separately and also to the ``full" stitched $HK_1$ spectrum. 

The top panels of Fig.~\ref{fig:sptype} and Fig.~\ref{fig:sptypeboth} display $\chi^2$ values versus spectral type. The minimum $\chi^2$ occurs in the early T dwarf range for the $H$-band only spectrum. Meanwhile, the $K_1$ band and combined $HK_1$ comparisons both reach a minimum with an overall trend tending towards the L/T transition. Specific best matching stars are shown in Table~\ref{tab:bestmatch}. From this analysis we qualitatively adopt a spectral type of T1$\pm$2 noting the large scatter in the goodness-of-fit statistic between filter fits and fact that brown dwarfs with similar spectral features can have different physical properties anyhow.

\begin{figure*}[!t]
\begin{center}
\includegraphics[height=5in]{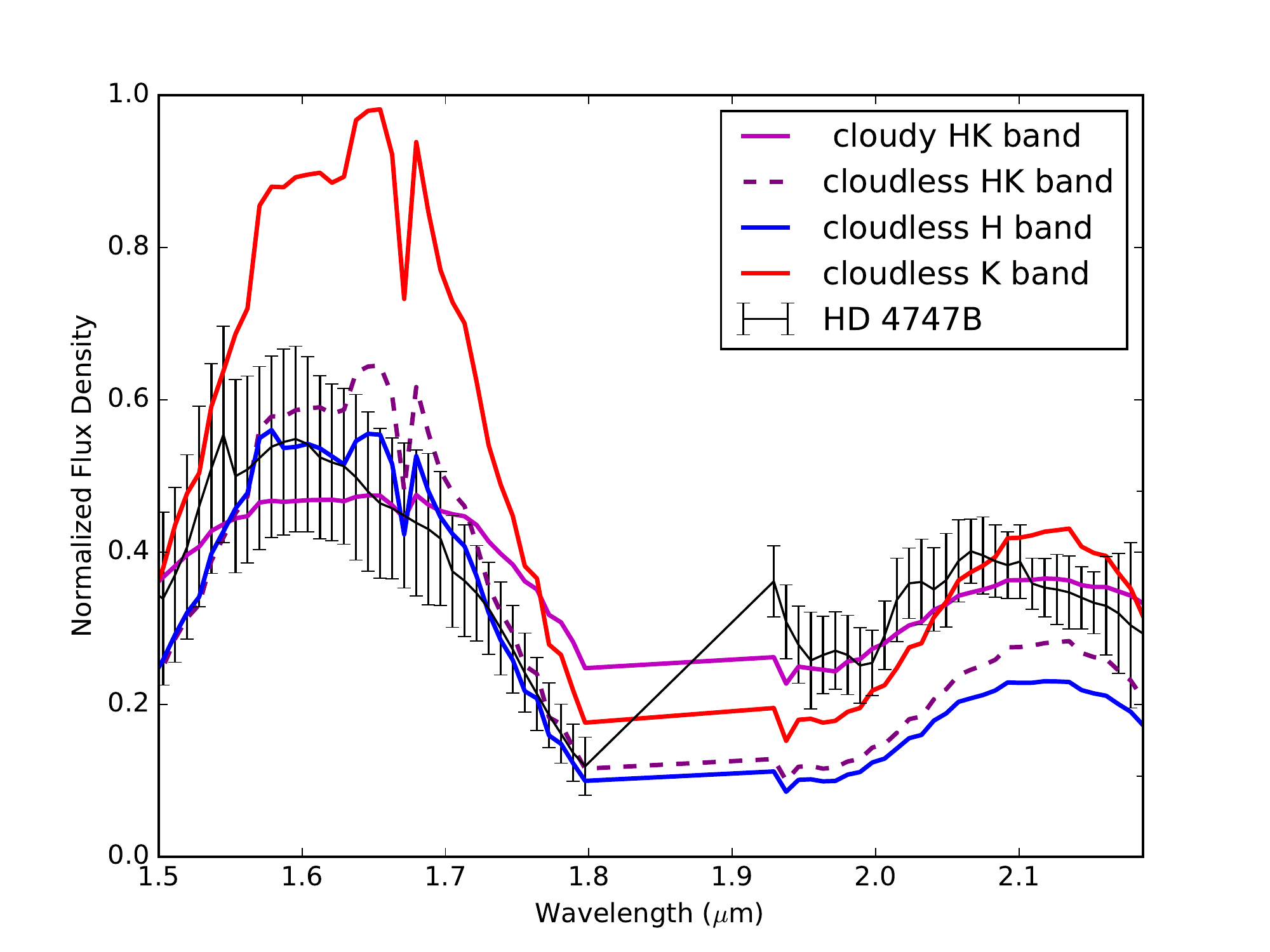}  
\caption{Model fitting results using the \citet{saumon_marley_08} models for individual filters ($H$, $K_1$) compared to a complete $HK_1$ fit. Cloudless models can reproduce only one filter spectrum at a time. The presence of clouds corroborates the result that HD~4747~B is an L/T transition object.}
\end{center}
\end{figure*}


\subsection{Model Atmosphere Fits}\label{sec:atmospheres}
Brown dwarf spectral types generally correlate with temperature. However, in order to infer physical properties, it is necessary to use methods complementary to the NIR spectral typing analysis performed above. We use the \citet{saumon_marley_08} atmosphere model grid of synthetic spectra to fit HD 4747~B and infer its temperature. The models are designed to reproduce the L/T transition covering the temperature range needed for this application \citep{saumon_marley_08}. They also incorporate an adjustable sedimentation factor ($f_{\rm sed}$) and vertical mixing parameter ($k_{zz}$) to describe cloud clearing that nominally occurs during this transition. The models range from $T=900 -1600$ K in increments of $\Delta T = 100$ K evaluated at surface gravities $\log(g)=$4.0, 4.5, 5.0, 5.5 (cgs units), sedimentation factor of $f_{\rm sed}=0$ (cloudless) or $f_{\rm sed}=2.0$ (cloudy), and eddy diffusion coefficient $k_{zz}$ of either 0 (chemical equilibrium) or 4.0 (not in chemical equilibrium). 

\begin{table}[ht]
\centerline{
\tablinesep=0.1ex
\begin{tabular}{lccc}
\hline
\hline
Parameter                  & $H$                               & $K_1$ & $HK_1$  \\
\hline \hline
$T_{\rm eff}$ [K]     &  $1509^{+159}_{-141}$    &   $1393^{+138}_{-150}$       &  $1407^{+134}_{-140}$     \\ 
$\log (g)$ [cm/s$^2$]  &  $5.2^{+0.5}_{-0.5}$  &  $4.5^{+0.4}_{-0.5}$  & $5.2^{+0.5}_{-0.6}$     \\ 
$f_{\rm sed}$              & $0$                                   & $0$                                  & $1.91^{+0.18}_{-0.18}$   \\
$k_{\rm zz}$               & $0$                                   & $0$                                  & $0$         
\\                          
\hline \hline
\end{tabular}}
\caption{HD~4747~B model fitting results for $H$-band only, $K_1$-band only, and the combined $HK_1$ bands using the \citet{saumon_marley_08} grids.}
\label{tab:theory_fits}
\end{table} 

The fitting routine begins similarly to the spectral typing analysis above by binning and trimming the models to match the wavelength range and resolution of the spectrum of HD~4747~B. The model spectra are normalized to the companion spectrum via a normalization constant (Cushing et al. 2008). A $\chi^2$ value is calculated for every binned model compared to the companion spectrum to determine a starting point on the model grid that results in a small burn-in phase for the next step in the Bayesian procedure. Probability distributions are generated using a Markov Chain Monte Carlo (MCMC) approach (Foreman-Mackey et al. 2013). The MCMC routine begins at the minimum $\chi^2$ value parameter space point and linearly interpolates between model spectra flux.


We perform the fitting routine for the individual $H$ and $K_1$ bands separately as well as the combined $HK_1$ spectrum (Fig. 3). The $H$ and $K$ band best fit models were normalized to the individual band; the full spectrum was multiplied by the same normalization constant and over-plotted on the companion spectrum for comparison. We also plot the combined $H$ best fit model at the same $T_{eff}$, log(g), and $k_{zz}$ parameters with an $f_{\rm sed}$ of 0 (cloudless). 

We find that cloudless models are able to adequately fit only one filter data set at a time. Cloudy models do a much better job replicating the spectrum of HD~4747~B across the $H$ and $K_1$ bands, but discrepancies in overall infrared color remain. Similar results have been obtained for other brown dwarfs that show comparable spectra but distinct offsets in integrated flux (color) \citep{leggett_02,konopacky_16}. This feature is commonly cited as a ``secondary parameter" problem that involves missing physics in the models related to metallicity, gravity, cloud and dust properties, unresolved binaries, the influence that electron conduction has on cooling, and other items that models may inherently assume such as spherical geometry, hydrostatic equilibrium, nonmagnetic, non-rotating, etc. \citep{chabrier_00,saumon_marley_08}. 

In any case, model fitting of our GPI spectrum suggests that HD~4747~B is a cloudy brown dwarf. Table~\ref{tab:theory_fits} summarizes the results. Correlation matrices between parameters are shown in Figure~5. HD~4747~B shows an effective temperature that is consistent for each wavelength range to within $\Delta T \approx 100$ K. From this analysis, we adopt an effective temperature of $T_{\rm eff}=1450 \pm 50$ K. This temperature is consistent with an L/T spectral type \citep{stephens_09,filippazzo_15}.

Model fits yield only relatively weak constraints on the companion surface gravity. The $K_1$-band fit suggests the lowest surface gravity ($\log g = 4.5_{-0.5}^{+0.4}$ dex), whereas the $H$-band fit ($\log g = 5.2_{-0.5}^{+0.5}$ dex) and full $HK_1$ spectrum ($\log g = 5.2_{-0.6}^{+0.5}$ dex) prefer a higher gravity (Table~\ref{tab:theory_fits}). In these models, $k_{zz}=0$ while $f_{\rm sed}=0$ for the individual bands and $f_{\rm sed}=1.9$ for the joint $HK_1$ fit. 


Since surface gravity is not always well constrained via model fitting of low resolution spectra, we also infer $\log(g)$ from the companion dynamical mass (see $\S$\ref{sec:dynamical_mass}) and radius estimated from evolutionary models. We use the independent gyrochronological age of the system, $\tau = 3.3^{+2.3}_{-1.9}$ Gyr, which was calculated in \citet{crepp_16}, based on the host star $B-V$ value and estimated rotation period; this value includes scatter in the activity-rotation empirical relations from \citet{mamajek_hillenbrand_08}. We next use the \citet{saumon_marley_08} models to determine an age-dependent radius, finding $R=0.089^{+0.006}_{-0.005}R_{\odot}$. The radius in turn yields an estimate for the surface gravity, $\log(g)=5.33^{+0.03}_{-0.08}$, which is based on the dynamical mass but is still model-dependent. Nevertheless, it offers a self-consistency check when comparing surface gravity with effective temperature, isochrones, and other parameters as is done in $\S$\ref{sec:evol_models}. For comparison, we find $R=0.087^{+0.005}_{-0.007}R_{\odot}$ and $\log(g)=5.35^{+0.10}_{-0.08}$ using the \citet{baraffe_03} models. These surface gravity values are consistent with the spectral fitting results and also expected older age of the system, e.g. compared to young stellar clusters, moving groups, or associations.


\begin{figure*}[!t]
\begin{center}
\includegraphics[height=5.0in]{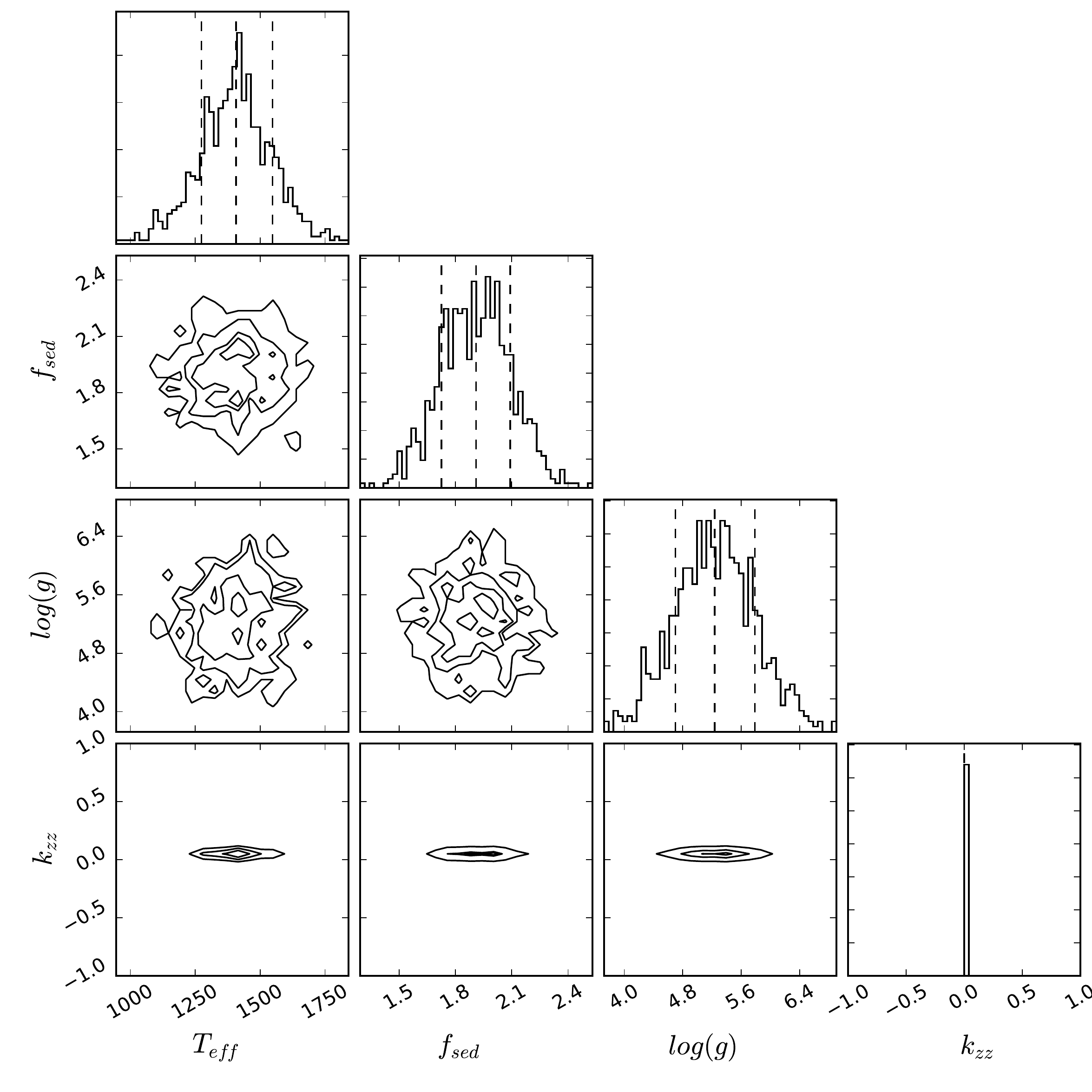} 
\caption{Combined $HK_1$ atmospheric model fitting results. Histograms are marginalized over other fitting parameters. See discussion.}
\end{center}
\label{fig:triangle}
\end{figure*}  




\subsection{Presence of Methane}\label{sec:methane}
Methane is a volatile compound whose presence as an absorption feature may be used to identify and classify brown dwarfs \citep{burgasser_99}. Methane absorption in the $H$-band influences the near-infrared color of substellar objects and can help to distinguish between L dwarfs and T dwarfs \citep{geballe_02}. To further assess our classification scheme, we analyze the H-band spectrum of HD~4747~B to search for methane by calculating a methane index, $S_{\rm CH_4}$,
\begin{equation}
S_{CH_4}=\frac{\int_{1.560}^{1.600} f_{\lambda} d\lambda }{\int_{1.635}^{1.675} f_{\lambda} d \lambda}
\end{equation}  
where $f_{\lambda}$ and $d\lambda$ values are derived from the GPI spectrum. This technique was also used for an H-band methane analysis of the benchmark brown dwarf HD~19467~B with the Project 1640 IFS \citep{crepp_15}. Brown dwarfs generally have methane indices $S_{CH_4} > 1$. We find that HD~4747~B has $S_{CH_4}=1.1\pm0.1$. Using values for the spectral classification of brown dwarfs from \citet{geballe_02}, we infer a spectral type of L9.5-T2 (Fig.~6). This range is consistent with our model fitting analysis and further supports the interpretation that HD~4747~B is an L/T transition object.

\begin{figure*}[!t]\label{fig:methane_index}
\begin{center}
\includegraphics[height=3.2in]{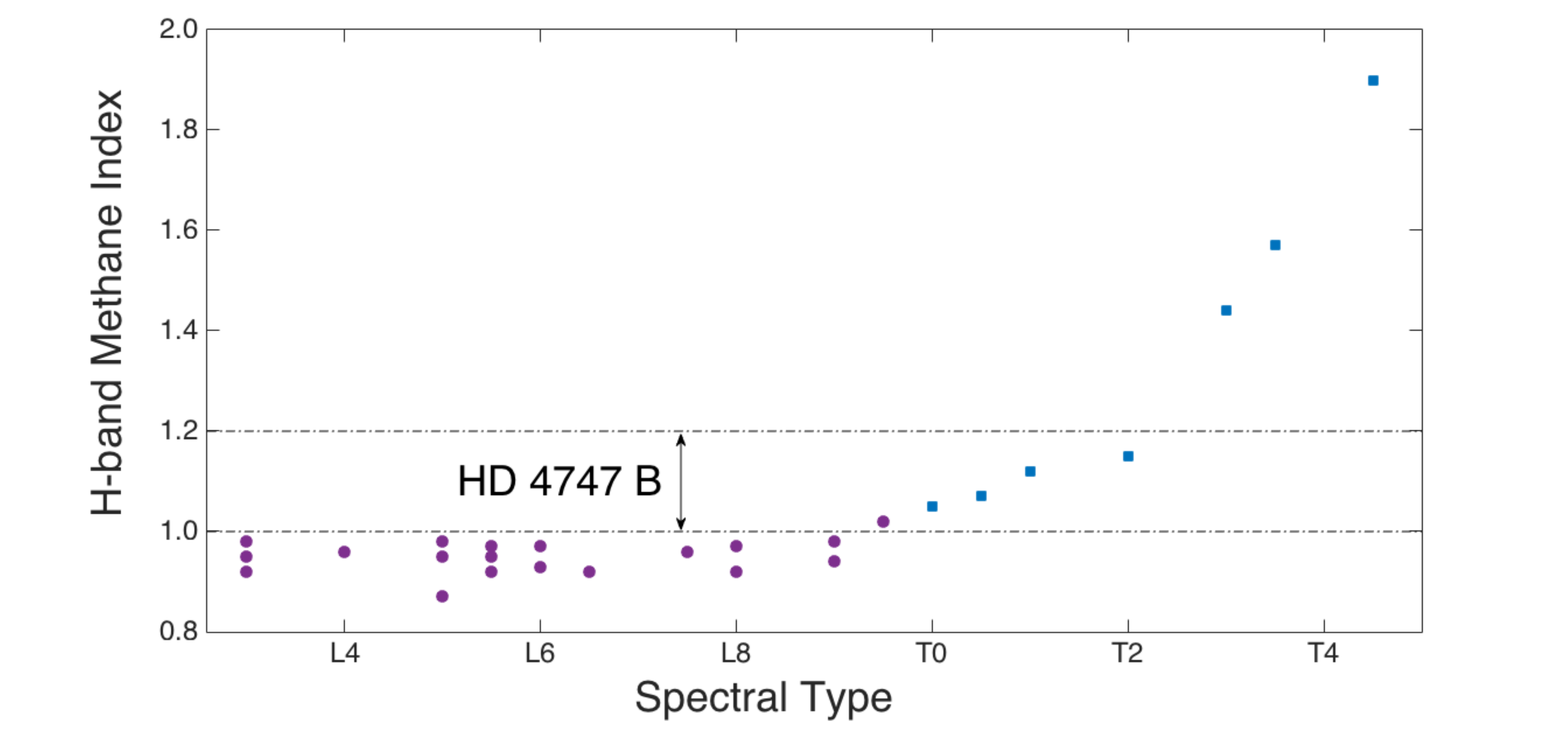} 
\caption{H-band methane indices for brown dwarfs from \citet{geballe_02} plotted as a function of spectral type. HD~4747~B has an H-band methane index of $S_{CH_4}=1.1\pm0.1$ as indicated by the horizontal lines. This range is consistent with a spectral type of L9.5-T2.}
\end{center}
\end{figure*} 


\begin{table*}[!t]
\centerline{
\begin{tabular}{cccccc}
\hline
\hline
JD-2,450,000     &  Instrument   &     Filter    &     $\rho$ [mas]      &    Position Angle [$^{\circ}$]     &   Proj. Sep. [AU]   \\
\hline
\hline     
6,942.8              &    NIRC2     &      $K_s$   &      $606.5 \pm 7.0$  &  $180.^{\circ}04 \pm 0.^{\circ}62$   &   $11.33 \pm 0.17$ \\ 
7,031.7              &    NIRC2      &      $K_s$   &     $606.6\pm 6.4$   &  $180.^{\circ}52 \pm 0.^{\circ}58$   &   $11.34 \pm 0.16$ \\
7,289.9              &    NIRC2      &      $L'$      &      $604 \pm 7$       &   $184.^{\circ}9 \pm 0.^{\circ}9$      &      $11.3 \pm 0.2$   \\
\hline   
7,380.6              &  GPI          &     $K_1$     &     $585\pm14$    &    $185.2^{\circ}\pm0.3^{\circ}$       &    $10.93\pm0.29$ \\
7,381.5              &  GPI          &     $H$         &     $583\pm14$    &    $184.4^{\circ}\pm0.3^{\circ}$       &    $10.90\pm0.29$ \\
\hline
\hline
\end{tabular}}
\caption{Relative astrometric measurements of HD~4747~B including initial discovery data from NIRC2 (first three rows) and two new GPI observations (bottom two rows).}
\label{tab:astrometry}
\end{table*}

\subsection{Metallicity Considerations}\label{sec:metallicity}
At present, ``atmospheric models remain largely untested at non-solar metallicities" \citep{west_11}. This is due in large part to a lack of available benchmark objects, HD~114762~B representing the latest available spectral type (d/sdM9) at [Fe/H]=-0.7 \citep{bowler_09}. One of the goals of the TRENDS program is to detect benchmark companions with disparate metallicities. HD~4747 has a precisely measured metallicity of $\mbox{[Fe/H]}=-0.22 \pm 0.04$ dex. However, for all intents and purposes, this value is very nearly solar considering that metal content is expected to influence the emergent spectrum of brown dwarfs as a second order effect. In particular, only cloudless models are publicly available that explore metallicity. \citet{skemer_16} sample [M/H] at 0.0, 0.5, and 1.0 dex to study GJ 504 b, but custom models have not yet been developed for the L/T transition to our knowledge, so far only for transiting exoplanets \citep{morley_17}.
 
As a provisional test of the influence of metallicity on HD~4747~B, we consider the recently developed Sonora model grids \citep{marley_17}. Noting as a caveat that only cloudless models are currently available, we investigate qualitatively the spectral morphology and quantify the percentage difference between models sampled at solar metallicity and [Fe/H]$=-0.5$ dex. In the H-band, we find that the sub-solar model offers a slight better match to the GPI data though comparable to uncertainties. In the K$_1$-band, the peak flux is located at shorter wavelengths which arguably better reflects the spectral morphology of HD~4747~B, modulo the systematic flux offset noted previously. At most, the models differ by $\approx$15\% considering that they straddle the expected metallicity of HD~4747~B, presuming it is comprised of the same material as its parent star. In comparison, the cloudy and cloudless models presented in Figure 4 differ by 41\% on average across the H and K bands, indicating that the expected effect of subsolar metallicity on the spectrum of HD 4747~B is much less than that of its cloudy atmosphere. Theoretical work on the impact of metallicity on cloudy substellar objects is on-going and beyond the scope of this paper (Marley et al., in prep). 


\subsection{Astrometry}\label{sec:astrometry}
We measure the position of HD~4747~B relative to its parent star to constrain the system orbit and dynamical mass. Observed in coronagraphic mode, the primary star remains hidden behind the occulting spot so its position must be inferred using the off-axis satellite spots described in $\S$\ref{sec:data}. Following speckle suppression we measure the centroid of the companion (and each injected fake companions) in each wavelength channel to determine the angular separation, position angle, and their uncertainties. We use the plate-scale of $\theta=14.37\pm0.34$ mas pix$^{-1}$ from the GPI instrument calibration webpage based on data obtained by \citet{konopacky_14} for $\Theta$ Ori using the orbital information from Magellan AO \citep{close_13}. 

Table~\ref{tab:astrometry} shows astrometric results for $H$ and $K_1$ data respectively. Uncertainty in the plate scale is folded into the analysis by adding its contribution in quadrature with other uncertainties. We find that uncertainty in the position of HD~4747~B is dominated by speckles and residual atmospheric dispersion. Results from each filter are consistent with one another to within $1\sigma$ in angular separation and $2\sigma$ is position angle. The position of the companion with GPI are consistent with that of NIRC2 from the discovery article which we also show for reference \citep{crepp_16}. The companion appears to be moving in a counter-clockwise direction with an angular separation that is decreasing with time. Given that the GPI data was obtained only three months after the most recent NIRC2 data, we do not update the companion dynamical mass or orbit which has a period of approximately 38 years. Performing a statistical (MCMC) analysis of the system astrometry, we find that posterior distributions yield comparable uncertainties as the original discovery article with the new GPI measurements. 

\begin{sidewaysfigure*}[!t]\label{fig:params}
\begin{center}
\includegraphics[height=4.6in]{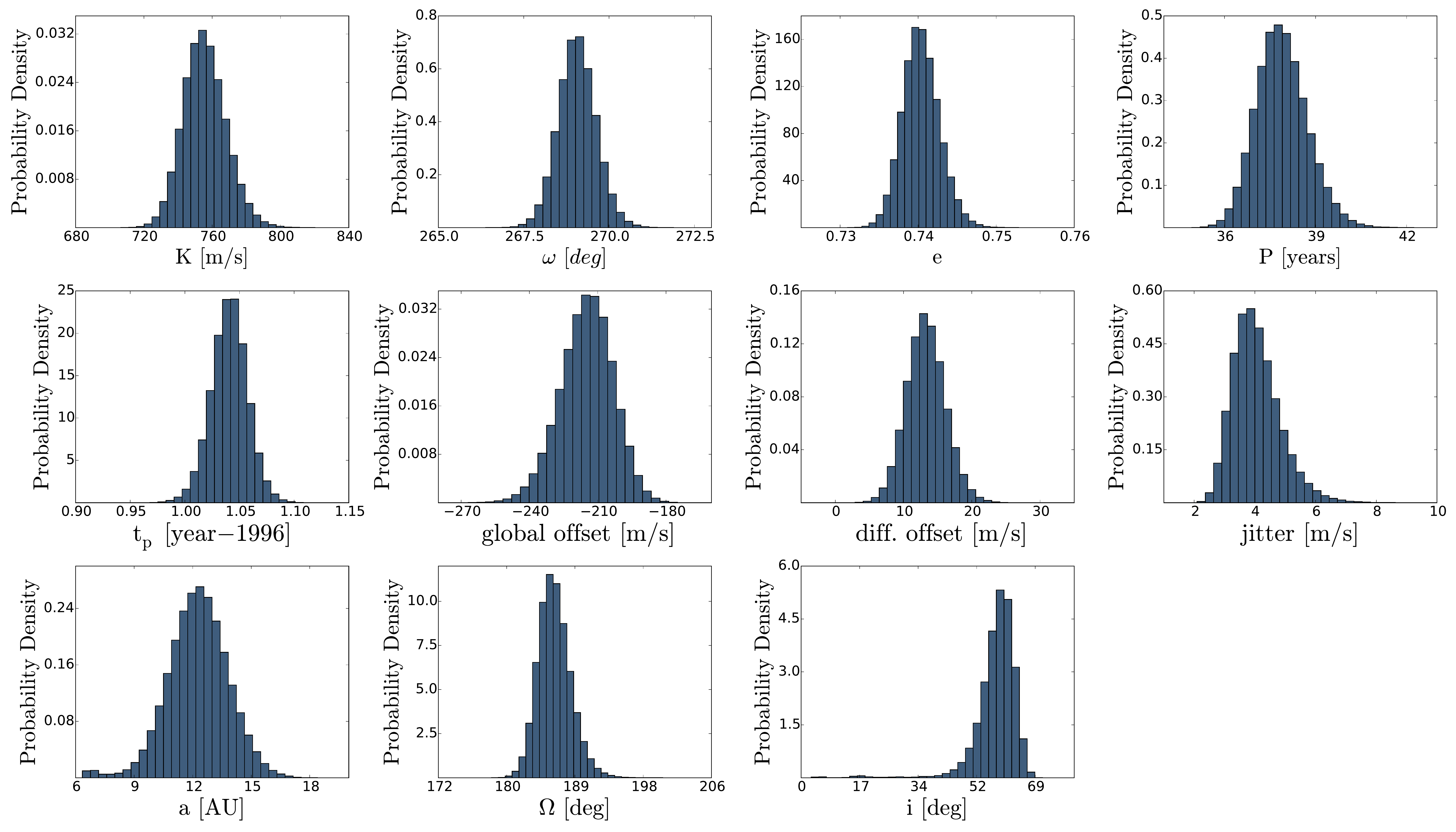} 
\caption{HD~4747~B orbital parameters posterior distribution.}
\end{center}
\end{sidewaysfigure*} 

\begin{figure*}[!t]\label{fig:orbit}
\begin{center}
\includegraphics[height=4.6in]{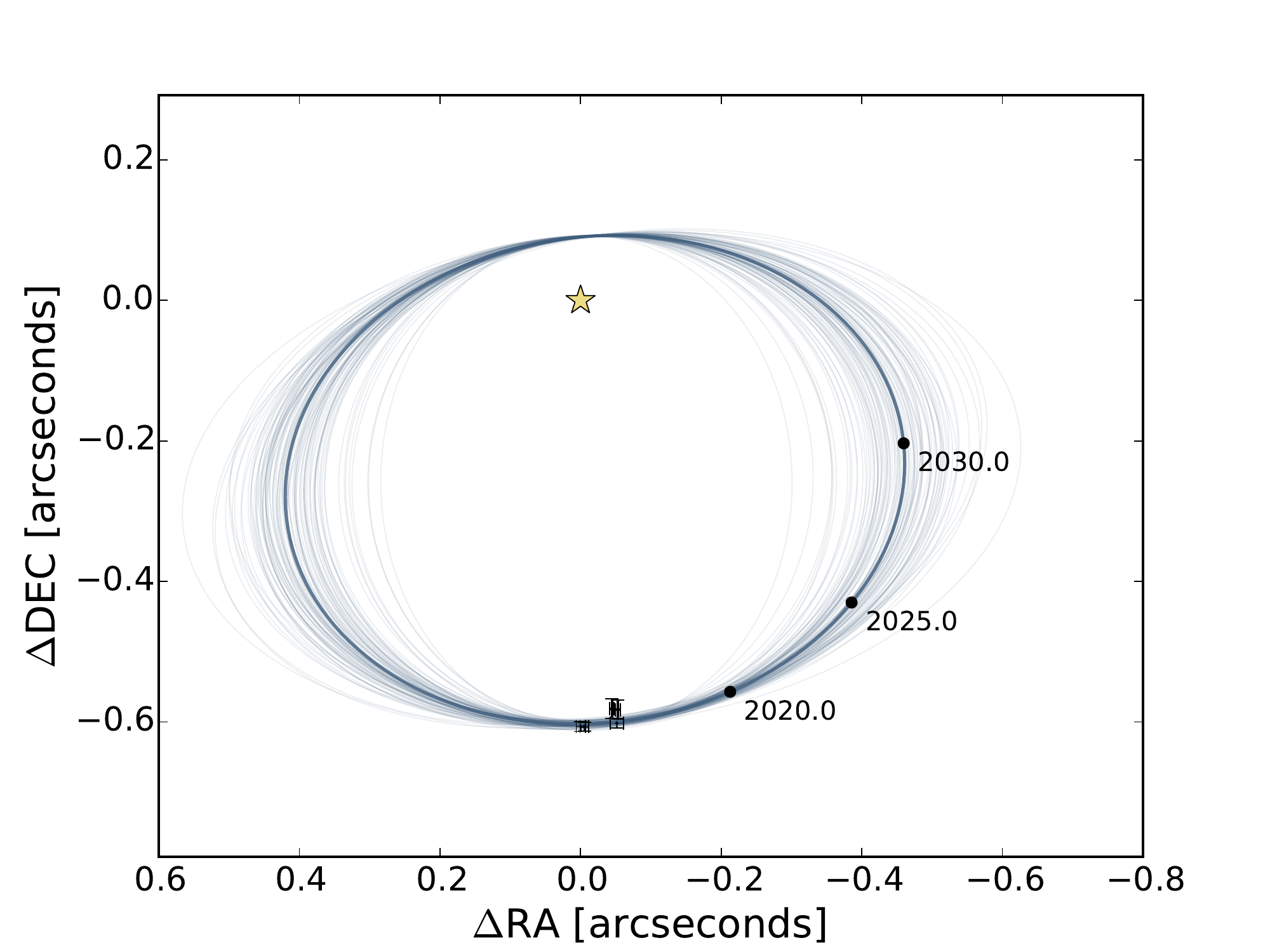} 
\caption{HD~4747~B sky-projected orbit.}
\end{center}
\end{figure*} 

\begin{figure*}[!t]\label{fig:mass}
\begin{center}
\includegraphics[height=4.6in]{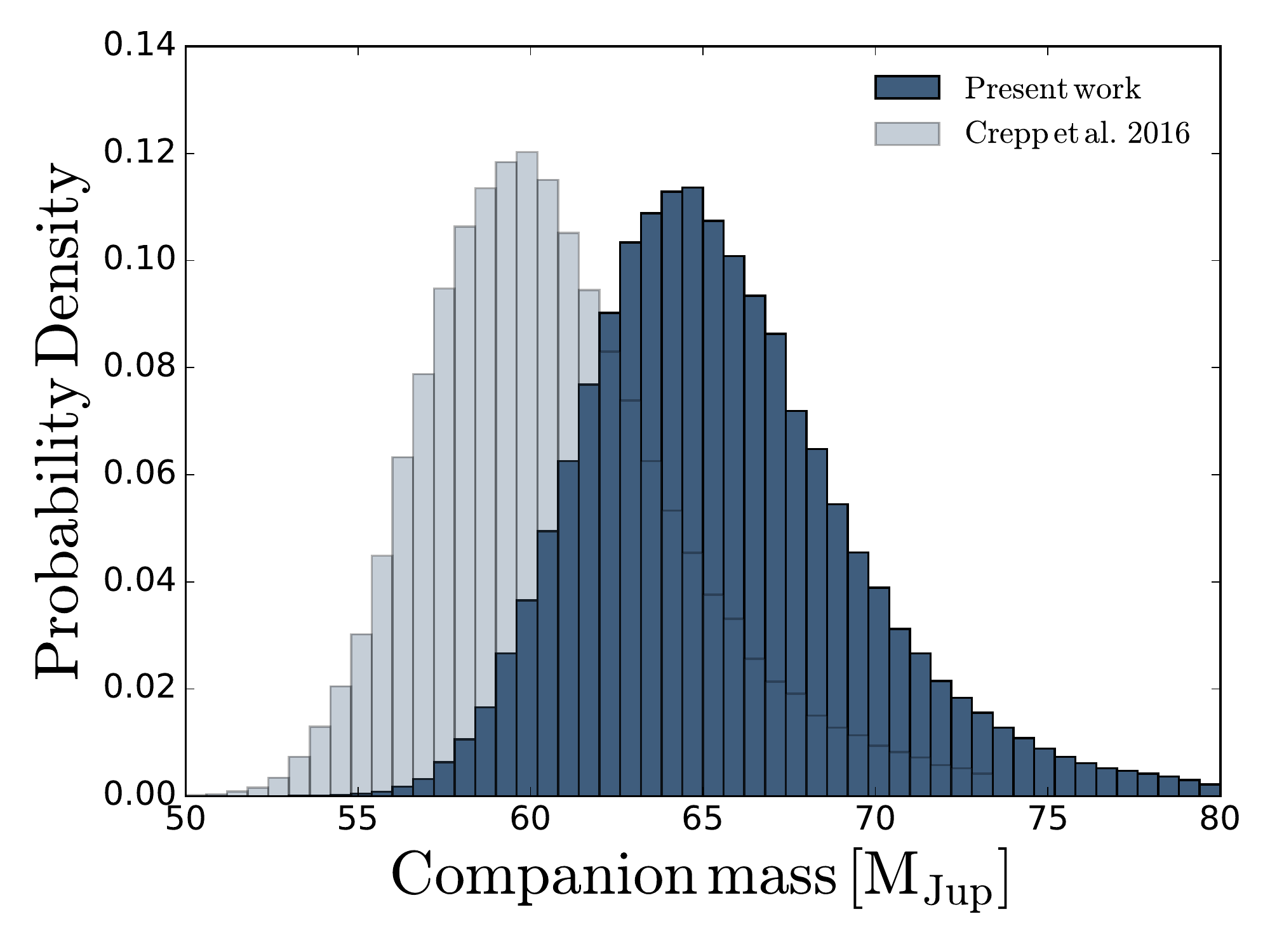} 
\caption{Mass posterior distributions before/after obtaining GPI astrometry data.}
\end{center}
\end{figure*} 

\subsection{Orbit and Dynamical Mass}\label{sec:dynamical_mass}
We update the dynamical analysis presented in the HD~4747~B discovery paper \citep{crepp_16}. We use the same methodology by jointly fitting the 18 years of stellar RV data with imaging astrometry from Table~\ref{tab:astrometry}. GPI measurements extend the astrometric time baseline by $\sim$92 days. This is small compared to the expected $P=38$ year orbit of the companion but nevertheless allows us to assess any systematic errors and perform consistency checks. Measurement uncertainties are comparable to the NIRC2 data set.

The python-based ``MCMC Hammer" routine is used within a Bayesian framework to calculate posterior distributions for orbital parameters ($K, P, e, \omega, \Omega, i, t_p$) where $K$ is the RV semi-amplitude, $P$ represents the orbital period, $e$ is eccentricity, $\omega$ is the longitude of periastron, $\Omega$ is the longitude of the ascending node, $i$ is inclination, and $t_p$ is time of periastron passage \citep{foreman_mackey_13}. Table~\ref{tab:orbitfit} shows results for the orbital parameter distributions including nuissance parameters such as the global RV offset, differential RV offset (resulting from two distinct HIRES instrument settings), and RV jitter value. 

We find that most parameters remain very similar to the original analysis with the exception of the orbital inclination and semi-major axis, which are now both smaller and with reduced uncertainties. Although it is still early to confidently extract orbital information from astrometry data given the limited baseline, the result of a lower inclination and semi-major axis was actually predicted by \citet{crepp_16} based on model-dependent estimates of the companion mass on the grounds of self-consistency. This result suggests that the GPI and NIRC2 measurements are compatible with one another at the $\sim10$ mas level. Figure~\ref{fig:orbit} shows the most recent sky-projected orbit trajectories consistent with both imaging and RV data sets with predictions for future epochs. 

\begin{table}[!t]
\centerline{
\begin{tabular}{lc}
\hline
\hline
\multicolumn{2}{c}{Orbital Fitting Results} \\
\hline
\hline
Parameter                        &     Value (68\%)                      \\   
\hline                                    
K  [m/s]                             &  $755.1^{+12.8}_{-11.6}$       \\ 
P [years]                           &  $37.85^{+0.87}_{-0.78}$       \\ 
e                                       &  $0.740^{+0.002}_{-0.002}$    \\ 
$\omega$ [$^{\circ}$]       &  $269.3^{+0.6}_{-0.6}$            \\ 
$t_p$ [year]                      &  $1997.04^{+0.02}_{-0.02}$    \\  
$i$ [$^{\circ}$]                  &  $58.5^{+4.0}_{-5.2}$               \\   
$\Omega$ [$^{\circ}$]      &  $186.2^{+2.3}_{-1.7}$             \\   
$a$ [AU]                           &  $12.3^{+1.5}_{-1.5}$               \\  
\hline
Global offset [m/s]             & 	$-214.4^{+10.9}_{-11.9}$  \\ 
Diff. offset [m/s]      	        &   $13.2^{+2.8}_{-2.7}$          \\ 
Jitter [m/s]                 	        &   $4.0^{+0.9}_{-0.7}$                 \\  
\hline
mass [$M_{Jup}$]             &    $65.3^{+4.4}_{-3.3}$           \\
\hline
\hline
\end{tabular}}
\caption{Orbital parameters and dynamical mass of HD~4747~B.}
\label{tab:orbitfit}
\end{table}

Finally, we use values directly from the MCMC chain parameters for $K$, $i$, $e$, and $P$ to solve for the companion dynamical mass. To accommodate for uncertainty in the mass of the host star, we draw from a normal distribution centered on the spectroscopically determined mass from the discovery paper, $M_*=0.82\pm0.04M_{\odot}$ \citep{crepp_16}.\footnote{We note that the stellar mass was determined using the iterative version of Spectroscopy Made Easy (SME) that analyzes high resolution spectra along with Yonsei Yale isochrones \citep{valenti_fischer_05}. Uncertainty in the companion mass enters through the Doppler RV equation as $M_*^{2/3}$ power and currently does not dominate the width of the dynamical mass posterior distribution, which is more-so governed by the time baseline of astrometry. To assess how systematics in the model-dependent host star mass impact the derived companion mass, we consider as an example a 10\% change from $M_*=0.82M_{\odot}$ to $M_*=0.90M_{\odot}$. We find that the companion mass increases by $\Delta m=3.9M_{Jup}$, comparable to the current uncertainty of several Jupiter masses.} We numerically calculate the companion mass from each MCMC iteration to generate a posterior distribution. While the star mass may contain small, model-dependent systematic errors, this method ensures that uncertainty in the companion mass scales as $m \sim M_*^{2/3}$ making this contribution comparable to uncertainties that result from minimal astrometry orbital coverage. 

Figure~\ref{fig:mass} shows results for the companion mass posterior before and after including GPI measurements. We find that the most likely (mode) companion mass has increased slightly since the discovery data, from to $60.2 \pm 3.3M_{\rm Jup}$ to $65.3^{+4.4}_{-3.3}M_{\rm Jup}$, a consequence of the smaller inclination. The posterior distribution is still asymmetric and actually slightly broader than before indicating that additional measurements are needed to further constrain the companion mass. Nevertheless, this information is sufficient to estimate the companion surface gravity, as was done in $\S$\ref{sec:atmospheres}, and compare to evolutionary models. 

\subsection{Comparison to Evolutionary Models}\label{sec:evol_models}
\begin{figure*}[!t]
\begin{center}
\includegraphics[height=5.0in]{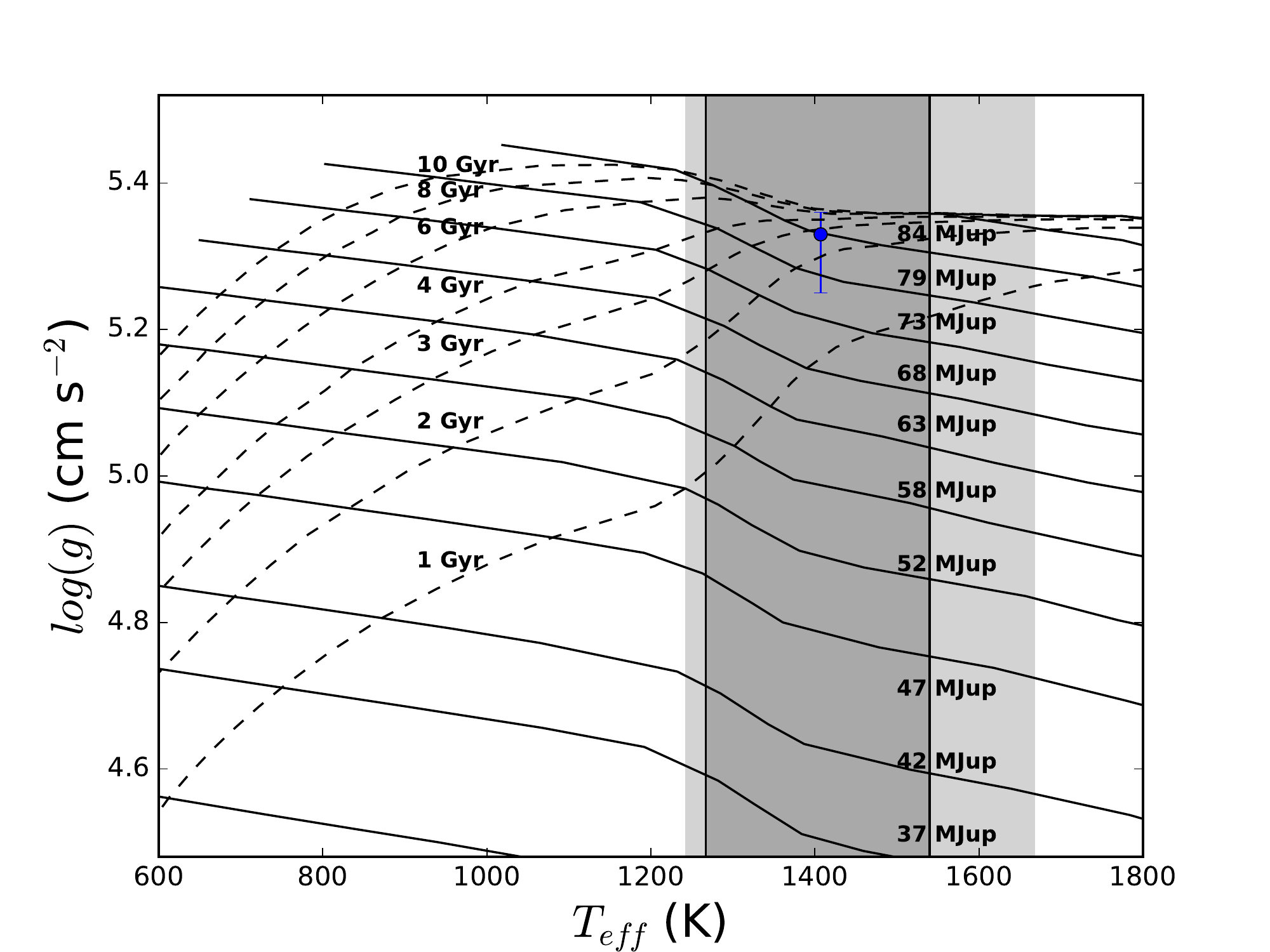} 
\caption{Surface gravity and effective temperature estimates for HD~4747~B. Overlaid are evolutionary model contours from \citet{saumon_marley_08}. We find a $2\sigma$ tension between the various constraints derived from the dynamical mass, surface gravity, effective temperature, and age (see text for discussion).}
\end{center}
\label{fig:logg}
\end{figure*}

We compare results for the effective temperature and surface gravity of HD~4747~B to brown dwarf evolutionary models. Contours from the \citet{saumon_marley_08} grids are shown in Fig.~\ref{fig:logg}. The range of viable $T_{\rm eff}$ values is indicated by shaded regions based on the results from spectral fitting (Table~\ref{tab:theory_fits}) where dark gray corresponds to the $HK_1$ best fit. The vertical extent of the plot is restricted to the range of $\log$ (g) values from Table~\ref{tab:theory_fits}. In Figure~\ref{fig:logg}, this corresponds to the entire plot. Given that $\log$ (g) is only loosely constrained from spectral fitting, we also display the surface gravity uncertainty from the companion dynamical mass and model-dependent radius. As discussed in $\S$\ref{sec:atmospheres}, the radius is estimated from the gyrochronology age using the same brown dwarf evolutionary models for self-consistency.

Comparing contours and intersections from the \citet{saumon_marley_08} grid to the allowable effective temperature and $\log$ (g) range, we find a tension ($\sim2\sigma$) between the companion dynamical mass $m=65.3^{+4.4}_{-3.3}M_{Jup}$ and that predicted from evolutionary models. At an age of $\tau = 3.3^{+2.3}_{-1.9}$ Gyr, the mass, effective temperature, and surface gravity are not coincident (Fig.~9). In order to rectify this inconsistency, either: (i) the dynamical mass would have to increase by $\approx 3 M_{\rm Jup}$; (ii) the effective temperature would have to decrease by $\approx 100$ K; or (iii) the companion age would need to be lowered by several hundred Myr which increases the companion radius and decreases $\log$ (g) by $\approx0.1$ dex. Each of these options can be tested with either further astrometric monitoring, higher resolution spectra obtained over a wider passband, or direct measurements of the host star radius using interferometry respectively. In the latter case, the age is inferred from the parent star based on its position in an HR diagram \citep{valenti_05,crepp_12a}. This is a model-dependent procedure that should be compared to gyrochronology in search of discrepancies  \citep{brown_14,maxted_15}. While age estimates are often suspect, we find that increasing the companion age only exacerbates the inconsistency while decreasing the companion age would reduce the surface gravity and thus offer better overlap between HD~4747~B and evolutionary model contours. 



\section{SUMMARY}\label{sec:summary}

We have used the GPI coronagraph and IFS to obtain the first near-infrared spectra of the benchmark brown dwarf discovered to orbit HD~4747. A substellar companion with a mass, age, and metallicity that have been determined independent of the light that it emits, HD~4747~B offers unique diagnostic capabilities for performing precise astrophysical tests of theoretical evolutionary and spectral models. The Sun-like host star has a precise parallax ($\pi=50.37 \pm 0.46$ mas) and the companion orbit ($a=12.3^{+1.5}_{-1.5}$ au) is amenable to dynamical studies on timescales of only several years, since more than 18 years of precise stellar RV data has already been recorded \citep{crepp_16}.   

Comparing HD~4747~B's $H$ and $K_1$ spectrum to other brown dwarfs, we find that the object is near the L/T transition with best-fitting spectral types of T1$\pm2$ ($\S$\ref{sec:sptype}), consistent with the original estimate from \citet{crepp_16} of $\approx$L8 based the companion mid-infrared $K-L'=1.34 \pm 0.46$ color and absolute magnitude $M_{K_s} = 13.00 \pm 0.14$. Theoretical models developed by \citep{saumon_marley_08} support this result in two ways: (i) temperature estimates are consistent with the L/T transition around $T=1450\pm50$ K; and (ii) neither cloudy nor entirely cloudless models adequately fit the data ($\S$\ref{sec:atmospheres}). This result if further corroborated by the detection of methane ($S_{CH_4}=1.1\pm0.1$) whose abundance empirically points to a similar spectral type of L9.5-T2 ($\S$\ref{sec:methane}). 

We find that theoretical models can well-reproduce the spectrum of HD~4747~B in individual filters, either $H$ or $K_1$, but systematic discrepancies arise when attempting to match the companion color when considering both filters simultaneously ($\S$\ref{sec:atmospheres}). This result is perhaps not surprising given that fully self-consistent cloud models are not yet available/tested for evolution across the L/T boundary, and the infrared colors of brown dwarfs with similar spectral features are known to show significant scatter \citep{leggett_02,konopacky_14}. The NIR color dispersion is related to secondary parameters involving composition and cloud parameters. Electron conduction may also play a role in the cooling of more massive and older brown dwarfs like HD~4747~B \citep{saumon_marley_08}. This effect is generally not incorporated into the physical model, nor is the impact of non-spherical geometry, rotation, magnetic fields, or processes such as precipitation fall-out, entrainment, and mixing. 

Although the new GPI astrometric measurements were obtained only $\approx$100-400 days after the NIRC2 discovery data, we find that the companion semi-major axis has decreased by 25\% ($\S$\ref{sec:astrometry}). This result was predicted by \citet{crepp_16} based on available star and companion mass estimates through a self-consistency analysis. Given the relatively short orbital period of the directly imaged companion, $P\approx38$ yr, even marginal improvements in the direct imaging astrometric baseline can impact the semi-major axis and inclination posterior at this early stage. Additional measurements from NIRC2, GPI, and other high-contrast systems are warranted because multi-instrument comparisons can introduce systematic errors at the level of several mas. Continued follow-up will further refine the companion dynamical mass and orbit ($\S$\ref{sec:dynamical_mass}). As noticed with other massive substellar companions located in the ``brown dwarf dessert", the high eccentricity of the companion ($e=0.74$) has yet to be explained \citep{crepp_12a,jones_17}.

There are a number of ways to further investigate the HD~4747~B companion. In particular, our GPI data includes only a fraction of the wavelength coverage offered by the instrument. Measurements at shorter wavelengths (YJ) would provide a lever arm for studying the companion spectrum, helping to further constrain atmospheric properties and study color anomalies. Measurements in the $K_2$ filter would provide access to another methane feature complementing the analysis performed in $\S$\ref{sec:methane}. Broad and simultaneous wavelength coverage using a lower resolution mode, such as that offered by the recently commissioned CHARIS instrument at Subaru, would help to substantiate any color anomalies, not just with HD~4747~B but other brown dwarfs as well \citep{groff_16}. 


Low resolution observations will not however help to constrain the surface gravity. We find a slight ($2\sigma$) tension when comparing the companion surface gravity, dynamical mass, effective temperature, and gyrochronological age to evolutionary models ($\S$\ref{sec:evol_models}). Either several of these parameter estimates require refinement, or shortcomings in the theoretical models are limiting comparisons made at this level of precision (see above discussion). Metallicity must also be mentioned as a caveat that can influence spectral morphology ($\S$\ref{sec:metallicity}), but cloudy models that explore metal content at the L/T transition are only now being developed \citep{marley_17}. In an effort to further understand the system age, we have recently obtained measurements of the host star radius from the CHARA Array. These results will be presented in a separate article (Wood et al., in prep.). Finally, GPI's polarimetry mode should be used to study condensate dust grains and near-infrared scattering properties \citep{jensen-clem_16}.


\section{ACKNOWLEDGEMENTS}
The authors thank the referee for a helpful and timely report. We thank Mark Marley for correspondence in studying the effects of metallicity and sharing early results from the Sonora models. Data presented in this article has leveraged a significant amount of work and computer programming developed by the GPI team and Gemini Observatory. In particular, efforts by Fredrik Rantakyro, Rene Rutten, and Pascale Hibon from Gemini Observatory were essential in securing this data set. The TRENDS high-contrast imaging program is supported in part by NASA Origins grant NNX13AB03G. JC acknowledges support from the NASA Early Career Fellowship. We are grateful for the vision and support of the Potenziani family and the Wolfe family.

\end{document}